\documentclass[12pt]{iopart}

\usepackage{graphicx}
\usepackage{epsfig}
\usepackage[english]{babel}
\usepackage{amssymb}

\newcommand{\be}{\begin{equation}}
\newcommand{\bea}{\begin{eqnarray}}
\newcommand{\eea}{\end{eqnarray}}
\newcommand{\ee}{\end{equation}}

\def\C{{\cal C}}

\def\one{\ensuremath{\hbox{$\mathrm I$\kern-.6em$\mathrm 1$}}}

\def\*{\star}

\def\[{\left[}
\def\]{\right]}
\def\({\left(}
\def\){\right)}

\def\d{\partial}

\def\2pi{\hbox{$2\pi i$}}

\def\dsl{\raise.15ex\hbox{/}\kern-.57em\partial}
\def\Dsl{\,\raise.15ex\hbox{/}\mkern-.13.5mu D}
\def\th{\theta}     
     
\def\al{\alpha}

\def\de{\delta}     

\def\2pi{\hbox{$2\pi i$}}
\font\numbers=cmss12
\font\upright=cmu10 scaled\magstep1
\def\stroke{\vrule height8pt width0.4pt depth-0.1pt}
\def\topfleck{\vrule height8pt width0.5pt depth-5.9pt}
\def\botfleck{\vrule height2pt width0.5pt depth0.1pt}
\def\Zmath{\vcenter{\hbox{\numbers\rlap{\rlap{Z}\kern
0.8pt\topfleck}\kern 2.2pt
                   \rlap Z\kern 6pt\botfleck\kern 1pt}}}
\def\Qmath{\vcenter{\hbox{\upright\rlap{\rlap{Q}\kern
                   3.8pt\stroke}\phantom{Q}}}}
\def\Nmath{\vcenter{\hbox{\upright\rlap{I}\kern 1.7pt N}}}
\def\Cmath{\vcenter{\hbox{\upright\rlap{\rlap{C}\kern
                   3.8pt\stroke}\phantom{C}}}}
\def\Rmath{\vcenter{\hbox{\upright\rlap{I}\kern 1.7pt R}}}
\def\Z{\ifmmode\Zmath\else$\Zmath$\fi}
\def\Q{\ifmmode\Qmath\else$\Qmath$\fi}
\def\N{\ifmmode\Nmath\else$\Nmath$\fi}
\def\C{\ifmmode\Cmath\else$\Cmath$\fi}
\def\R{\ifmmode\Rmath\else$\Rmath$\fi}

\begin{document}

\letter{Particle Content of the Nonlinear Sigma Model with $\theta$-Term: 
a Lattice Model Investigation}

\author{L Campos Venuti$^1$, C Degli Esposti Boschi$^2$, E Ercolessi$^{1,2,3}$, 
\\F Ortolani$^{1,2,3}$, G Morandi$^{1,2,3}$, S Pasini$^{1,3}$ and M Roncaglia$^{1,2,3}$}
\address{$^1$ Physics Department, University of Bologna, 6/2 v.le B.~Pichat, I-40127, Bologna, Italy.}
\address{$^2$ INFM, Research Unit of Bologna, 6/2 v.le B.~Pichat, I-40127, Bologna, Italy.}
\address{$^3$ INFN, Section of Bologna, 6/2 v.le B.~Pichat, I-40127, Bologna, Italy.}

\date{\today}

\begin{abstract}
Using new as well as known results on dimerized quantum spin chains with
frustration, we are able to infer some properties on the low-energy 
spectrum of the O(3) Nonlinear Sigma Model with a
topological $\th$-term. In particular, for sufficiently strong
coupling, we find a range of values of $\th$ where a singlet bound
state is stable under the triplet continuum. On the basis of these results,
we propose a new renormalization group flow diagram for the
Nonlinear Sigma Model with $\th$-term.
\end{abstract}


\ead{\mailto{roncaglia@bo.infn.it}}

\vspace{1 cm}
\noindent
A large class of 1D quantum spin models maps onto the O(3) Nonlinear Sigma Model (NL$\sigma$M),
described by a vector field ${\bf n}$, with the constraint $|{\bf n}|=1$. The action is written as
\be
\label{nlsm}
S=\frac{1}{2g}\int {\rm d}\tau {\rm d}x \(\frac{1}{v} |\d_\tau {\bf n}|^2+ v |\d_x {\bf n}|^2 \)+i\th T\, ,
\ee
where $g$ is the coupling constant and the topological charge $T$ is an integer number given by
\be
T=\frac{1}{4 \pi}\int {\rm d}\tau {\rm d}x \ \ {\bf n}\cdot \(\d_\tau {\bf n} \times \d_x {\bf n}\)\, .
\ee
The angle $\th$ determines the phase-shift contribution of a given field configuration
in every topological sector.

It is well known that in the case of antiferromagnetic Heisenberg chains, the coefficient of the topological
term is $\th=2\pi S$ \cite{HAue}.
So, the topological term is irrelevant for integer-spin chains, which show a dynamically generated spin gap
and exponentially decaying correlation functions.  
For half-odd integer spins, we have $\th=\pi$ and the model is gapless
with nonuniversal critical exponents, corresponding in the infrared limit to the SU(2)$_{k=1}$
Wess-Zumino-Novikov-Witten (WZNW) model, a $c=1$ conformal field theory \cite{Af,GNT}.

In a recent Letter \cite{CM}, the O(3) NL$\sigma$M for generic values of $\th$ 
close to $\pi$ has been approached using form factor perturbation theory on a related double-frequency sine-Gordon model.
As soon as one moves away from $\th=\pi$ towards $\th=0$, parity (${\bf n}\to{\bf - n}$) is broken and
the spinons get confined, forming a stable triplet of massive particles, with mass $M_{\rm t}$.
The authors of reference \cite{CM} have obtained an indication that above the triplet there exists also a singlet state,
with mass $M_{\rm s}$, which remains stable at least in a small interval of $\th$ around $\pi$. 
As it is known from the exact solution \cite{Zamo}, at the limiting value $\th=0$
there is no trace of the singlet state.
Hence, they conjectured the existence of a critical value $\th_{\rm c}$, below which the singlet state
becomes a resonance, that lies above the two-triplet continuum ($M_{\rm s}>2 M_{\rm t}$).
In this paper, we will discuss the same problem from the point of view of quantum spin chains.

The O(3) NL$\sigma$M with arbitrary values of $\th$ can be obtained starting from the
dimerized spin-1/2 chain with frustration:
\be
\label{Ham}
H= J\sum_i \Bigl[\(1+\delta(-1)^i\){\bf S}_i \cdot {\bf S}_{i+1}+\alpha\; {\bf S}_i \cdot {\bf S}_{i+2}\Bigr] ,
\ee
where $\delta$ is the parity-breaking coefficient responsible for the dimerization and $\al$
is a next-to-nearest neighbors (NNN) coupling constant, that for positive values induces frustration.
Besides its relevance to spin-Peierls \cite{exp} and spin-ladder \cite{WA} compounds, the model is interesting also
from a theoretical point of view as it contains two independent mechanisms for spin-gap formation.
At $\alpha = \alpha_{\rm c} \cong 0.241$ (\cite{BKJ} and references therein) there is a transition 
into a spontaneously dimerized ground state, for $\alpha > \alpha_{\rm c}$ a gap appears in the 
spectrum even with $\delta=0$ and the elementary excitations are massive spinons.
For $\al$ close to the critical value and in absence of dimerization,
we can use bosonization to obtain the sine-Gordon (SG) action:
\begin{equation}
{\cal A}^{\rm eff}_\pi = \int  {\rm d} \tau {\rm d} x \(\frac{1}{2} \partial_\mu
\varphi \partial^\mu \varphi + \gamma \cos{\beta \varphi}\)\, ,
\label{GpMT}
\end{equation}
where both $(8 \pi-\beta^2)$ and $\gamma$ are proportional to $(\al-\al_{\rm c})$ and lie on the marginally irrelevant 
branch of the separating curve in the Kosterlitz-Thouless RG-flux diagram, as required by the SU(2) invariance \cite{FGN}.
Switching on the dimerization, $\delta \neq 0$, gives rise to another term proportional to
$\delta \sin \frac{\beta}{2} \varphi$, which is strongly relevant and is responsible for the opening of a gap
$M_{\rm t}$ for $\alpha < \alpha_{\rm c}$.
The corresponding elementary excitation is a magnon triplet \cite{Af,GNT}.
The model with both $\delta$ and $\gamma$ different from zero is known as 
the double-frequency sine-Gordon (DSG) model.

Using the usual Haldane ansatz around the
Ne\'el classical solution, and taking the continuum limit after
integrating out the fast modes \cite{HAue}, the model Hamiltonian (\ref{Ham}) maps onto
the O(3) NL$\sigma$M \cite{RS}, with 
\be \label{param}
g=\frac{2}{S\sqrt{1-4\al -\de^2}}\, ,  \qquad \th=2\pi S(1-\de) \, ;
\ee
and $v=4J/g$. The domain of validity
of the classical solution is $4\al +\de^2<1$ \cite{RS}. The equations 
(\ref{param}) are not to be intended as a strict correspondence
between the bare lattice couplings ($\delta,\al$) and the
NL$\sigma$M, since we have partially lost the correct parameter
renormalization during the continuum limit procedure. 
The Haldane mapping is supposed to work well when the correlation lenght, $\xi$,
is much larger than the lattice spacing, $a$, which is surely realized for large $S$.
However, there are many examples in the literature showing that
the continuum approach is reliable down to $S=1/2$ for ladders \cite{PRL97} and to $S=1$
for chains \cite{PRB00}, examples for which $\xi/a$ turns out to be of the order of few units.
For our purposes we can adopt equations (\ref{param}) as approximate relations between 
microscopic and continuum parameters, in order to infer the particle content of the NL$\sigma$M 
from the low-energy spectrum of the spin chain.

\begin{figure}
\begin{center}
  \epsfxsize=10cm
  \epsfbox{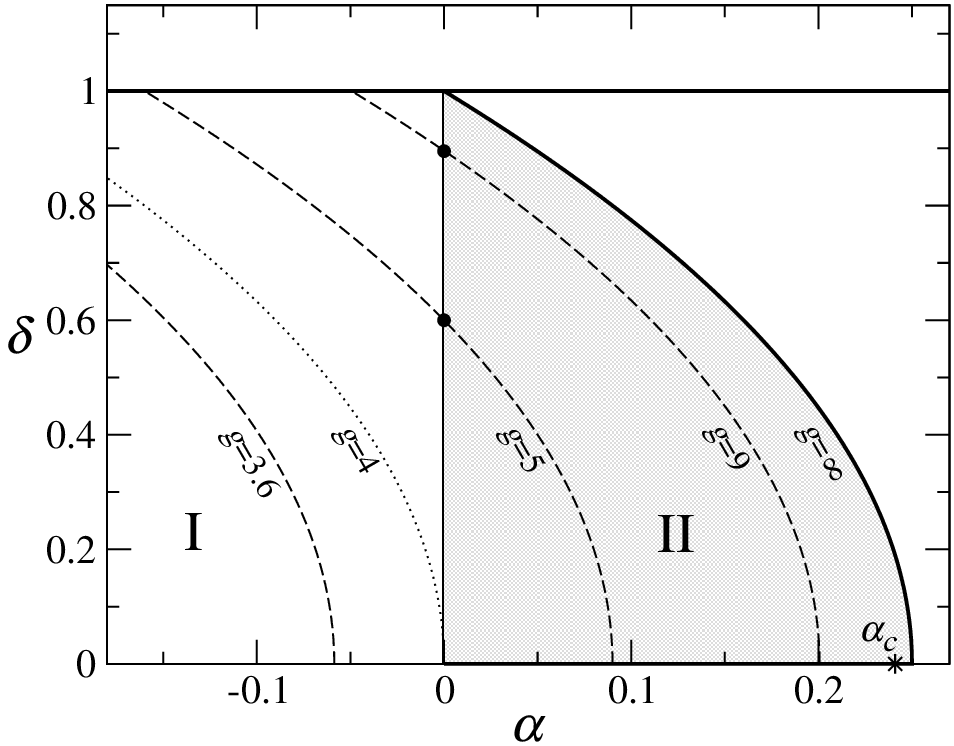}
  \caption{The Ne\'el phase in the dimerized $S=1/2$ chain with frustration.
In the shaded region (II) the singlet state is stable, while in region (I) is not.
The curves given in equation(\ref{parabola_al_del}) are plotted for several values of $g$ (dashed).
For reference, we have indicated the spontaneous dimerization point $\al_{\rm c}\cong 0.241$.}
  \label{fig1}
\end{center}
\end{figure}

Since in the model (\ref{nlsm}) the parameter $g$ essentially sets the energy scale,
we investigate the behaviour of the lowest levels when $\th$ is varied and $g$
is held fixed.
This can be realized in the lattice model by varying $\th$ via $\delta$
while moving $\alpha$ on the parabolas:
\begin{equation}
\alpha(\delta) = \frac{1}{4}\left( 1 - \delta^2 - \frac{16}{g^2} \right)\, ,
\label{parabola_al_del}
\end{equation}
which are displayed in figure \ref{fig1}, for several values of $g$.
Note that for $g > 4$ the curves span positive (region (II)) as well as negative (region (I)) 
values of $\alpha$, while for $g < 4$ they lie entirely in region (I). 
This introduces the need for investigating the
model (\ref{Ham}) also in region (I) where $\al<0$. 
To our knowledge, region (I) has never been studied in the literature, 
perhaps because it was not considered relevant for experiments.

One of the most important peculiarities of the spin model (\ref{Ham})
is that regions (I) and (II) mark two different low-energy physics. 
In region (I) the first excited states form a magnon
triplet, labeled by the quantum numbers $S_{\rm TOT}^z=-1,0,1$.
Besides the triplet, in region (II) there appears also a
singlet state. This fact was observed numerically by exact
diagonalization in reference \cite{BKJ}, where the mass ratio $R=M_{\rm s}/M_{\rm t}$ 
was computed for $0\leq\al<0.5$, which includes our region (II).
It was found that in this range the singlet is always stable, that is $R <2$,
the ratio $R$ changing monotonically with $\al$ and reaching
the two-particle continuum ($R=2$) at $\al=0$.
Moreover, it was observed that, as long as $\alpha <\alpha_{\rm c}$ the functions $R(\alpha)$
obtained at different values of $\delta$ seem to collapse onto a single universal curve,
showing that $R$ is almost independent from $\delta$.
The value $R \cong \sqrt{3}$ is found only at $\alpha=\alpha_{\rm c}$,
consistently with the fact that in this case the only nonvanishing potential term 
in the DSG model is $\sin{\sqrt{2 \pi} \varphi}$.
Therefore, on the basis of the results of \cite{BKJ} and keeping in mind equations
(\ref{parabola_al_del}) and (\ref{param}) we are led to
identify the critical $\theta$, at a given $g$, with:
\begin{equation}
\theta_{\rm c}=\pi(1-\delta_{\rm c})\, ,\qquad
\delta_{\rm c}=\sqrt{1-\frac{16}{g^2}}\, ,\qquad g>4 \, ;
\label{th_c}
\end{equation}
and $\th_{\rm c}=\pi$ for $g\leq 4$.

In order to support our identification of $\th_{\rm c}$ with equation (\ref{th_c}), we 
have to make sure that the singlet remains above the continuum in region (I),
where the NNN coupling is ferromagnetic and classically gives further contributions  
to the Ne\'el ordering created by the antiferromagnetic NN coupling.
Since this point was not addressed in the literature, and is at the core of our analysis,
 we have computed the mass ratio
for $\alpha < 0$ using the density-matrix renormalization group (DMRG) method,
which allows to investigate chains with several tenths of sites and a very good
numerical accuracy. 
As it will be discussed below, this is important because large system sizes
are necessary to obtain a good estimate of the singlet gap $M_{\rm s}$.
We used White's original finite-system algorithm \cite{Wh} with three iterations and the so-called thick-restart
Lanczos routine (with tolerance $10^{-10}$) to extract the ground state of the superblock as well as a given number
of excited states \cite{DEBO}. The number of optimized
DMRG states was fixed to $m=256$, and periodic boundary conditions have been used.
Although this choice in principle requires a larger $m$, its rationale is that
we expect a smaller finite-size contribution due to the absence of borders.
As a preliminary step, we have checked and reproduced some of the results of \cite{BKJ} 
in region (II). As regards $\alpha < 0$, we have selected four representative points
with the following results. Fixing $\delta=0.02$ we can follow the singlet state up to $L=120$ sites 
finding $R=2.06$ for $\alpha=-0.03$ and $R=2.07$ for $\alpha=-0.10$. With $\delta=0.10$, instead,
we estimate $R=1.96$ and $R=1.98$, for $\alpha=-0.03$ and $\alpha=-0.10$, respectively, on a range
of $L$ that does not exceed 80. The error on $R$ can be assumed to be a few percent, to the
extent that we considered the DMRG data to be reliable whenever the triplet degeneracy is
mantained within $0.5 \%$ and the fitted values of $M_{\rm t}$ and $M_{\rm s}$ also vary less than $0.5 \%$
when the maximal $L$ included in the fit is varied. Briefly, we can say that the singlet state
remains at threshold $R=2$ for $\alpha \le 0$. 

\begin{figure}
\begin{center}
  \epsfxsize=6.5cm
  \epsfysize=6.5cm
  \epsfbox{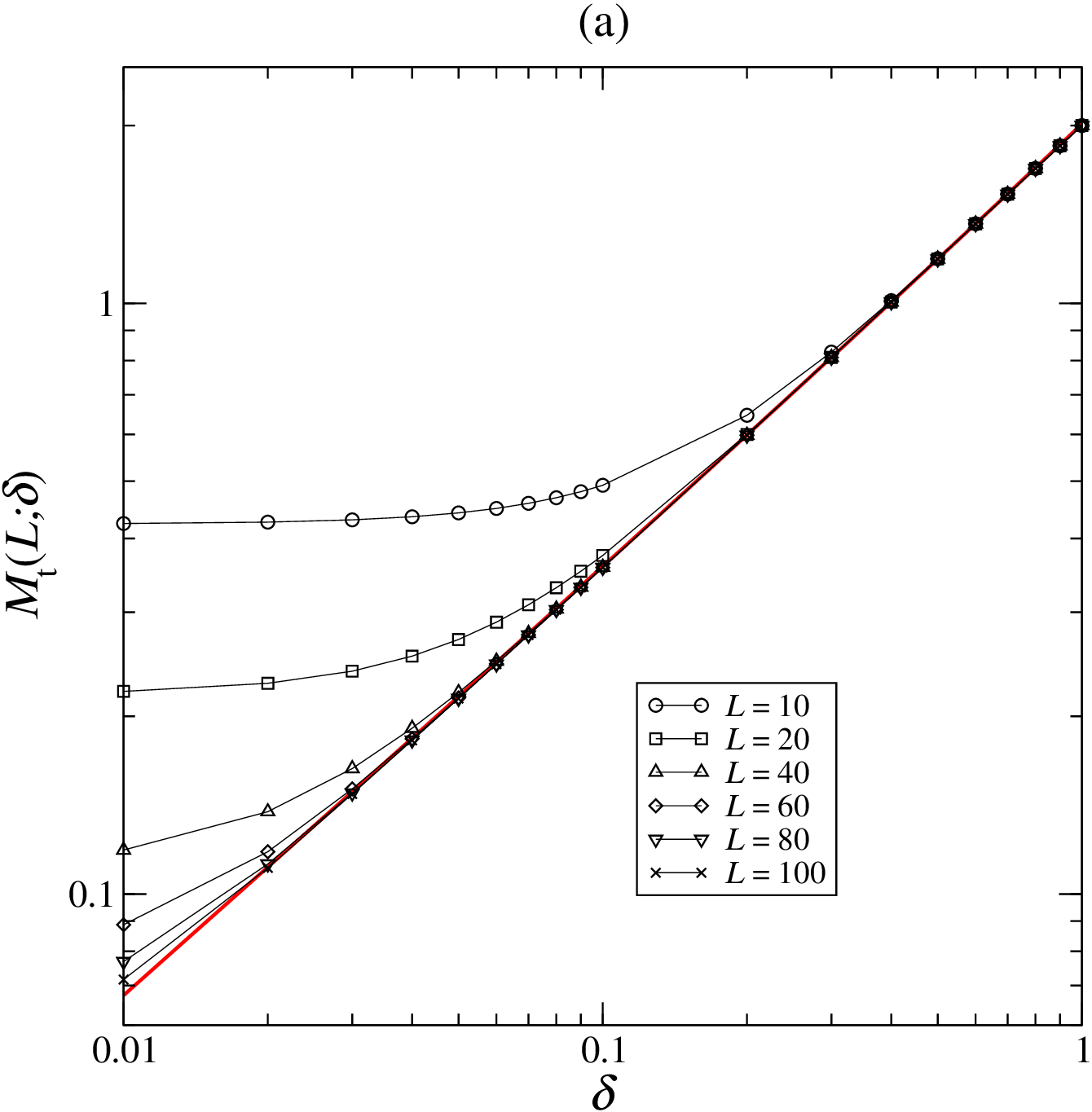}
  \epsfxsize=6.5cm
  \epsfysize=6.5cm
  \hspace{0.7cm}
  \epsfbox{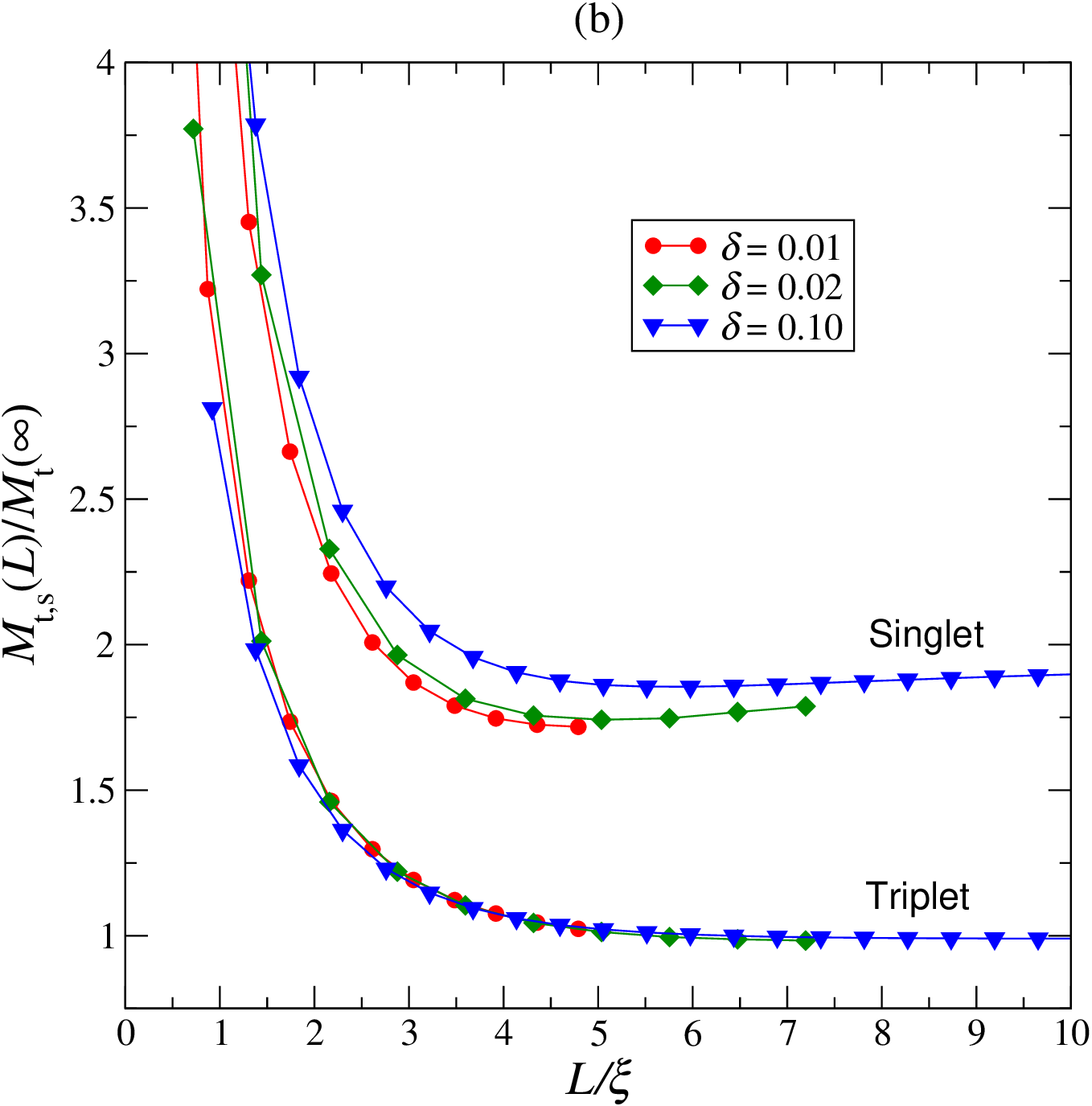}
  \caption{(a) Triplet gap $M_{\rm t}$ as a function of the dimerization $\delta$ for
           different lenghts at $\alpha = 0$. The solid line is the theoretical prediction
           equation (29) in reference \cite{Or}. 
           (b) Scaling plot of the singlet gap for the same parameters as above.
           The values of $M_{\rm t}(\infty)$ at different values of $\delta$
           are estimated directly using the theoretical curve in (a) and $\xi = v/M_{\rm t}$
           with $v = \pi/2$.}
  \label{fig2}
\end{center}
\end{figure}

In order to check in which regime our approach can provide accurate information on the spectrum
of the corresponding effective field theory, we have calculated 
the triplet gap $M_{\rm t}$ as a function of $\delta$ for the lattice model (\ref{Ham}) in the case $\alpha = 0$. 
Recently, Orignac \cite{Or} has provided detailed analytical predictions by using a renormalization group analysis that takes into 
account the logarithmic corrections due to the marginal term in the DSG model. From figure \ref{fig2}(a) we see that 
the DMRG data for the lattice model fit very well the theoretical curve calculated using the results in \cite{Or} 
also for large values of $\delta$, i.e. much beyond the small-gap regime. 
We notice that our data agree with the ones reported in \cite{PBD} and extend the analysis to a wider range. 

The infinite-size values of the gaps $M_{\rm t}$ and $M_{\rm s}$ used
to compute the ratio $R$ have been extrapolated by means of the fitting functions
given in \cite{BKJ}. While the finite-size triplet gap $M_{\rm t}(L)$ converges
monotonically from above, the analogous function $M_{\rm s}(L)$ for the singlet 
first exhibits a minimum at a characteristic length $L_{\rm min}$ and then 
saturates from below to its asymptotic value \cite{BKJ}.
To obtain an accurate estimate of $M_{\rm s}$ we considered only chains with at least 
$L_{\rm min}$ sites.
Since $L_{\rm min}$ gets larger as $\delta$ becomes smaller, we have to exclude
from our investigation values of $\delta \lesssim 0.01$, for which $L_{\rm min}$ exceeds
the maximum reliable size reachable with our DMRG. In previous exact diagonalization
studies \cite{BKJ}, $M_{\rm s}$ has been extrapolated on chains shorter than $L_{\rm min}$.
A possible justification for using a nonmonotonic fit on data that still do not display a minimum
lies in the existence of a scaling function that describes the finite-size 
effects at different $\delta$ (and fixed $\alpha$) once that the gaps and $L$ are
suitably rescaled by $M_{\rm t}$ and $\xi$ respectively. 
Figure \ref{fig2}(b) contains an example of such a scaling plot for $\alpha=0$ at 
three different values of $\delta$. It is seen that the data (especially the triplet)
tend to collapse onto two common curves showing that asymptotically $R$ does not
depend on $\delta$. A theoretical justification for these scaling functions is an open
problem which is currently under investigation.

For large $\delta$ all the excited states that we can target with the DMRG 
(up to ten in each sector of $S^z_{\rm TOT}$) are eventually exhausted by one-particle
excitations with energy $\sqrt{M_{\rm t}^2+v^2 q^2}$ and small momenta $q \propto 2 \pi/L$,
so that the singlet state is not detectable anymore. 
In summary, we have confidence on our DMRG data in an intermediate range
of $\delta$.

To study the singlet state for 
high dimerization values ($1/2\lesssim\delta\leq 1$), we have considered 
a completely different approach. 
A sophisticated study of the low-energy spectrum of the Hamiltonian (\ref{Ham}) was 
performed analytically in reference \cite{SKS}, treating the excitations as a gas of hard-core bosons, 
which are very dilute in the vicinity of the ``full-dimer'' point ($\delta=1,\al=0$).  
The authors have developed a detailed diagrammatic analysis that allows  
to calculate the single-magnon spectrum $\Omega_q$ with good accuracy. 
In the Ne\'el-like phases (I) and (II), the minimum of excitation lies at the centre of 
the Brillouin zone (BZ) $q=0$ (remember that we have doubled the unit cell). 
Once obtained the single-particle energies, it is possible 
to derive the Bethe-Salpeter equation for the singlet wave function $\psi_{\rm s}(q)$, as: 
\begin{equation}
[E_{\rm s}-\Omega_{q}-\Omega_{-q}]\psi_{\rm s}(q)= \int {\rm d} p\; M_{\rm s}(q,p)\psi_{\rm s}(p)\, ,
\label{bs}
\end{equation}
where $E_{\rm s}$ is the singlet-energy eigenvalue and $M_{\rm s}(q,p)$ 
is the scattering amplitude in the singlet channel (see reference \cite{SKS} for details). 
The singlet state here is viewed as a bound state formed by two elementary excitations 
with zero total momentum $Q=0$, getting contribution from various relative momenta $q$. 

\begin{figure}
\begin{center}
  \epsfxsize=9cm
  \epsfbox{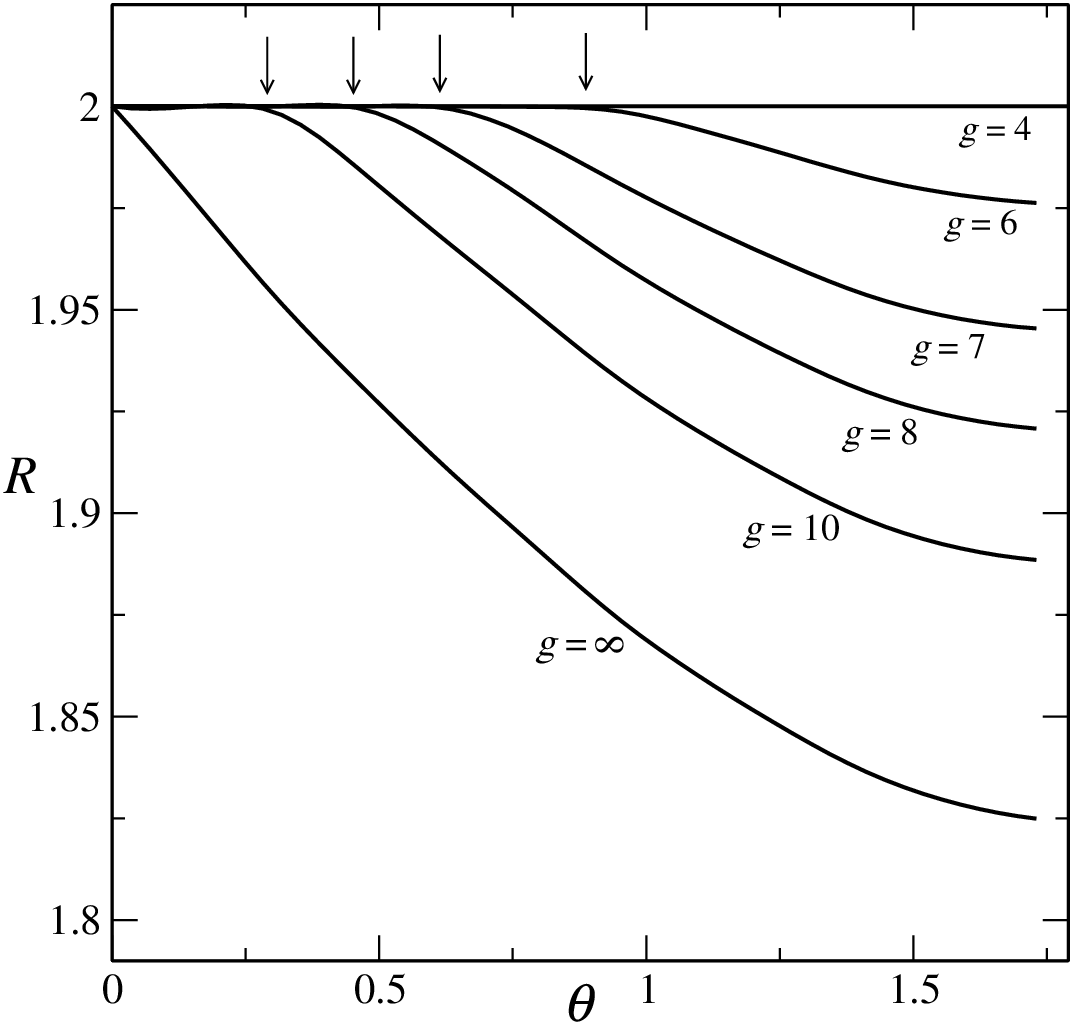}
  \caption{Singlet-triplet mass ratio $R$ as a function of the angle $\th$.
These values are calculated in the lattice model with the hard-boson method,
establishing via equation(\ref{param}) the connection with the NL$\sigma$M parameters $g$ and $\th$.
The arrows indicate the location of the critical value $\th_{\rm c}$ for various $g$.}
  \label{fig3}
\end{center}
\end{figure}

We have also calculated the singlet-triplet mass ratio $R$ for various 
values of $\al$ and $\delta\gtrsim 0.5$, adapted to our needs. 
We have found again that in region (I) the singlet state at rest stays 
at the continuum threshold, while in region (II) it becomes a stable particle. 
In figure \ref{fig3} we have plotted $R$ as a function of the angle $\th$ as predicted by the hard-core boson method,
establishing via equation (\ref{param}) the connection with the NL$\sigma$M parameters $g$ and $\th$.
This allows for a visual localization of the point where the singlet goes below the continuum $R=2$, 
identifying the critical value $\th_{\rm c}$, as conjectured in \cite{CM}. 
Note that as $g$ is increased $\th_{\rm c}$ shifts towards zero . 

One delicate issue concerns whether or not our observations are ``corrupted'' by
lattice effects that are not present in the continuum theory.
For instance, at $\delta=1$ the model (\ref{Ham}) can be thought as
a two-leg ladder with AF coupling $2J$ along the rungs and $\al J$ along the legs.
It is well-established that the NL$\sigma$M with $\th=0$ describes
the low energy sector of the even-legged ladder \cite{PRL97}.
For $\al>0$, the excitations are a magnon triplet with mass $M_{\rm t}$ at the
boundary of the BZ, $q=\pi/2$.
From the exact solution of the NL$\sigma$M \cite{Es}, it is known that the multi-particle continuum
sets in at $3M_{\rm t}$. However, calculations in the lattice for small $\al>0$
\cite{SK_JB} reveal that a singlet bound state appears at $q=\pi$ below a continuum with
threshold less than $3M_{\rm t}$, at variance with the field theoretical prediction. 
The reason for this discrepancy is that in the lattice, Lorentz invariance is manifestly broken.
In fact, at the boundary of the Brillouin zone, the continuum and the singlet
bound state get contribution from pairs of particles with momenta which are
far from $q=\pi/2$. Now, it is clear that a continuum theory that, for its nature, involves
a small range of momenta centred around $q=\pi$, cannot give rise to
such a continuum threshold and singlet binding energy.
On the contrary, in the present paper we have considered the region where
the classical Ne\'el phase is stable, giving rise to a magnon triplet
at $q=0$. Here the continuum starts at $2M_{\rm t}$ where it gets contribution
from single particle excitations around the centre of the BZ.
Moreover, we have also checked explicitly that $|\psi_{\rm s}(q)|^2$, the square of the solution of 
Bethe-Salpeter equation (\ref{bs}) for the singlet at $Q=0$, is sharply peaked around $q=0$. 
This fact tells us that the singlet at rest receives contributions mainly from single particles 
with small relative momentum. In other words, in the continuum limit where only 
long wavelength excitations are significant, the bound states are composed by slowly moving 
particles, thus belonging to the same theory.  
On the basis of this analysis we are confident that the information gained on the 
spectrum of the lattice model may be applied on the NL$\sigma$M with $\th$-term. 

Collecting results, some important facts seem to emerge, namely: 
\begin{itemize}
\item [i)] if $\th\leq \th_{\rm c}(g)$ the NL$\sigma$M with $\th$-term has only a triplet 
of elementary excitations and lies in the same universality class of the model 
without the topological term. 
\item [ii)] if $\th>\th_{\rm c}(g)$ the spectrum is enriched by the appearance of a stable 
singlet state with mass $M_{\rm s}=R M_{\rm t}$. 
The nonuniversal parameter $R$, labels different theories in the infrared limit, 
identifying an RG-flux invariant. 
This has been shown at least for $\th$ not too close to $\pi$ ($|1-\th/\pi|\lesssim 0.01$). This region is hard to investigate 
using the presently available methods, but in the sequel we will sketch some qualitative feature of the flux in the framework of the DSG. 
\item [iii)] the numerical observation that in the chain $R$ does not change with $\delta$ leads to 
the conclusion that all the NL$\sigma$M's corresponding to constant $\al$ flow towards the same point. 
So, it is possible to identify the RG-flux trajectories $\th(g)$: 
\be
\th(g)= \pi \(1\pm \sqrt{1-4\al-\frac{16}{g^2}}\)\, , 
\label{fluxR}
\ee
labeled by the flux invariant $\al$ (i.e. $R$). 
\end{itemize}

\begin{figure}
\begin{center}
  \epsfxsize=12cm
  \epsfbox{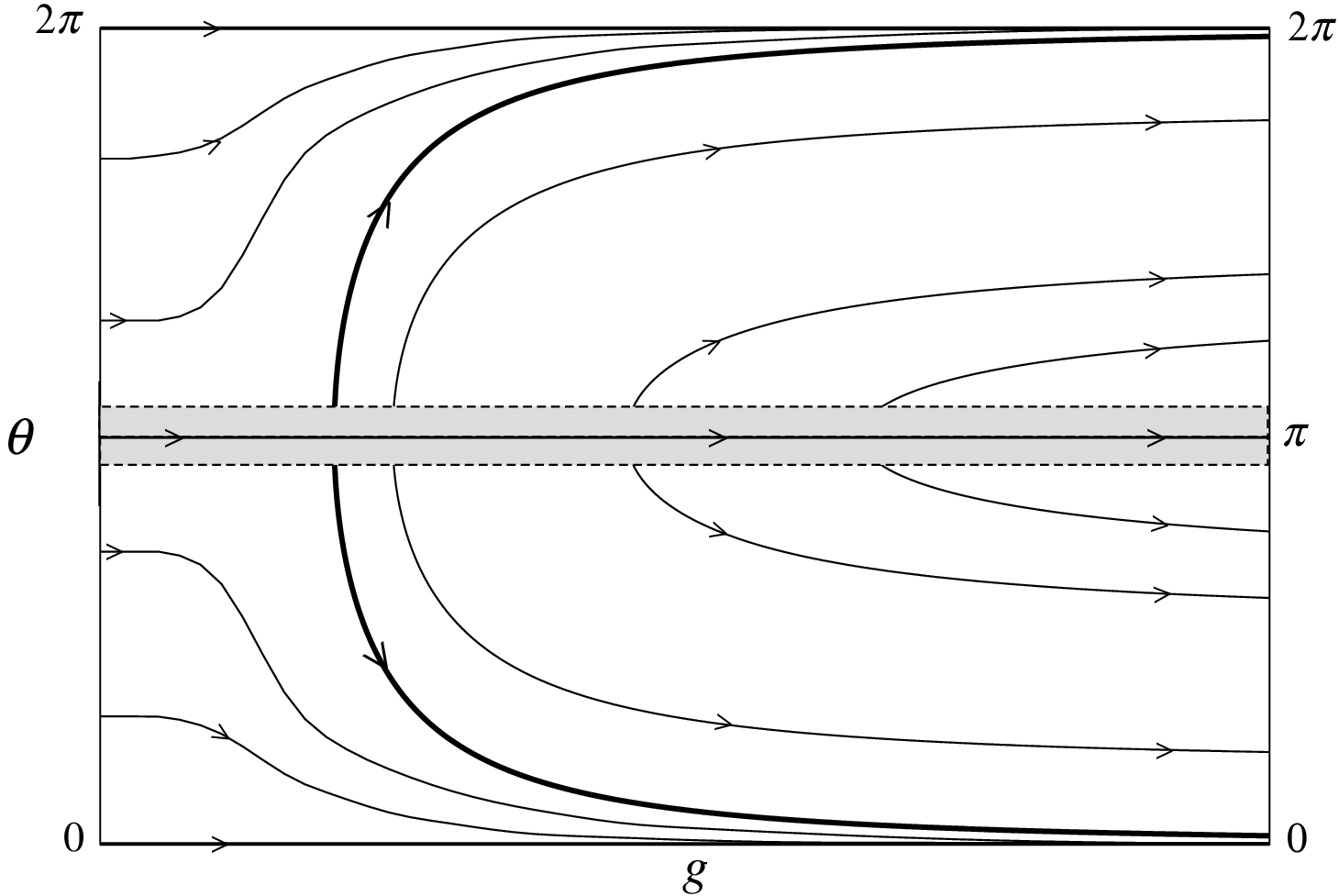}
  \caption{RG-flux diagram for the NL$\sigma$M with $\th$-term.
The thick line is separates the $\th=0$ regime (left) and
the nonuniversal regime (right) where the curves tend asymptotically
to $\th\neq 0$. With our arguments we cannot deduce the behaviour of the flux lines
inside a thin strip (shaded area, actually exaggerated) around the $\th=\pi$ line.}
  \label{fig4}
\end{center}
\end{figure}

All these considerations can be summarized in a proposal for the RG-flux diagram of the 
NL$\sigma$M with $\th$-term, that we have displayed in figure \ref{fig4}. 
We distinguish between two regimes separated by the curve given in equation (\ref{fluxR}) with $\al=0$ (thick line). 
On the left of this line, we have the $\th=0$ regime.  
At weak coupling $g$ renormalizes towards higher values while $\th$ does not renormalize. 
Hence the flux lines exit horizontally from $g=0$ and move to $\th=0(\textrm{mod}\, 2\pi)$ for strong coupling. 
Differently, on the right of the separating curve, the flux lines tend to different asymptotes 
$\th= \pi \(1\pm \sqrt{1-4\al}\)$, that identify the different theories labeled by $R$. 
Around the line $\th=\pi$ we have sketched a strip that covers a {\it quasi-critical} regime. 
We guess that all the flux lines that exit from this strip, actually should bend 
and orient parallel to the line $\th=\pi$ as they get closer to it.  
We can be convinced of this fact by looking at the DSG RG-flux \cite{YDX}. 
The parameter $\gamma\propto (\al-\al_{\rm c})$ flows marginally to zero, while $\delta$ is relevant. 
Starting with $\gamma$ and $\beta^2$ on the BKT separating curve and perturbing with a very small 
$\delta\ll \gamma$, it is possible to identify two regimes: a {\it quasi-critical} regime, where 
$\gamma\to 0$ faster than the growth of $\delta$; a {\it massive} regime where 
$\delta$ grows so rapidly that $\gamma$ may be regarded as a constant. 
On the basis of our calculation, we deal with parameters whose values fall in the second regime, 
while we cannot say much on the {\it quasi-critical} regime. 
In view of the fact that all the curves exit horizontally from $g=0$, only two scenarios are
possible, namely: i) either the two branches of the separating curve start at different points
on the $g=0$ axis or: ii) they originate from a fixed point $g=g^*$ of $O(1)$ on the $\th=\pi$ axis 
(where parity symmetry is spontaneously broken \cite{Af}) together with all 
the trajectories on the right of the separating curve.

In conclusion, we have obtained some interesting information on the 
low-energy spectrum of the NL$\sigma$M with $\theta$-term. 
We identify two regimes, a universal one where $\th$ flows always to $0(\textrm{mod}\, 2\pi)$
and a nonuniversal one where 
the infrared theories are labeled by a parameter $R$. This differs partially from what
suggested by Affleck in \cite{Af} due to the existence of the universal regime.
This leads to a 
new proposal for the RG-flux diagram of the NL$\sigma$M with $\theta$-term.   

\ack
The authors are grateful to G Mussardo and D Controzzi for interesting discussions
that have stimulated the subject of this paper. This work was partially funded by
the TMR network EUCLID (contract number: HPRN-CT-2002-00325), and the Italian MIUR
through COFIN projects (prot. n. 2002024522\_001 and 2003029498\_013).


\end{document}